# Dispersion independent long-haul photon counting OTDR

Bin Li,[2] Ruiming Zhang,[1] Yong Wang,[6] Hao Li,[6] Lixing You,[6] Zhonghua Ou,[3] Heng Zhou,[2] Yun Ling,[2] Yunxiang Wang,[3] Guangwei Deng,[1] You Wang,[1,4] Haizhi Song,[1,4] Kun Qiu,[2] and Qiang Zhou[1,3,5,*]

[1]*Institute of Fundamental and Frontier Sciences, University of Electronic Science and Technology of China, Chengdu 610054, PR China*
[2]*School of Information and Communication Engineering, University of Electronic Science and Technology of China, Chengdu 610054, PR China*
[3]*School of Optoelectronic Science and Engineering, University of Electronic Science and Technology of China, Chengdu 610054, PR China*
[4]*Southwest Institute of Technical Physics, Chengdu 610041, PR China*
[5]*CAS Key Laboratory of Quantum Information, University of Science and Technology of China, Hefei 230026, PR China*
[6]*State Key Laboratory of Functional Materials for Informatics, Shanghai Institute of Microsystem and Information Technology, Chinese Academy of Sciences, Shanghai 200050, China*
*\*Corresponding author: zhouqiang@uestc.edu.cn*

**Photon counting optical time-domain reflectometry (PC-OTDR) based on the single photon detection is an effective scheme to attain the high spatial resolution for optical fiber fault monitoring. Currently, due to the spatial resolution of PC-OTDR is proportional to the pulse width of a laser beam, short laser pulses are essential for the high spatial resolution. However, short laser pulses have a large bandwidth, which would be widened by the dispersion of fiber, thereby causing inevitable deterioration in spatial resolution, especially for long-haul fiber links. In this letter, we propose a scheme of dispersion independent PC-OTDR based on an infinite backscatter technique. Our experimental results -with more than 50 km long fiber - show that the spatial resolution of the PC-OTDR system is independent with the total dispersion of the fiber under test. Our method provides an avenue for developing the long-haul PC-OTDR with high performance.**

Optical time-domain reflectometry (OTDR) has been widely utilized as an indispensable tool for the nondestructive testing of optical fibers and systems [1, 2]. It has achieved a dynamic range of up to ~50.0 dB with the spatial resolution of around one kilometer [3]. However, due to the limitation of the detection bandwidth of photodetector (PD) [4], it is challenging to meet those high-end monitoring areas that require high spatial resolution, such as fiber-to-the-home (FTTH), airborne optical fiber network, and so on. With the development of single-photon detection technology, photon counting OTDR (PC-OTDR) based on a single photon detector (SPD) has received increasing attention because of its better spatial resolution and higher dynamic range [5-7].

In the PC-OTDR system, ignoring the time jitter of an SPD, the spatial resolution is defined by the pulse width of the laser [8]. The short-pulse laser has a large bandwidth, which would be broadened because of the dispersion of fiber, thereby causing inevitable deterioration in spatial resolution [9]. In order to prevent the laser pulse from being broadened in long distance applications, the bandwidth of pulsed laser should be as narrow as possible. Decreasing the bandwidth of the laser, however, will increase the optical pulse width, i.e. laser pulse obeying the Fourier transformation limitation, rendering poor spatial resolution in both short- and long-distance measurements [10]. Despite a spatial resolution of 6 cm has been achieved with long-haul fiber under test (FUT), the spatial resolution is hard to keep unchanged along the FUT due to the dispersion broadening of the laser pulse [11]. Moreover, a hybrid structure [12] consisting of two different performance PC-OTDR systems is employed to realize high-dynamic and high-resolution respectively. But the spatial resolution is also varying along the FUT, i.e. dependent on the dispersion property of the fiber. Thus, a dispersion independent long-haul PC-OTDR system is yet to be proved.

Fortunately, Deepak Devicharan *et al.* proposed an infinite backscatter technique for conventional OTDR technique [13]. It operates as follows: wide pulse of laser light, whose width is proportional to the length of FUT, is sent into the FUT periodically; as these pulses gradually illuminating the fiber, the lighten parts produce back-propagation optical signals due to Rayleigh scattering and Fresnel reflections; differential calculations of back-propagation optical signals measured at defined time periods result in an OTDR trace, known as the fiber loss profile. With the infinite backscatter technique, the spatial resolution is just defined by the measurement interval, i.e. the sampling rate of the analog to digital converter (ADC), not the pulse width itself, thus achieving high spatial resolution with high sampling rate. Compared with other novel dispersion-independent technique based on chaotic laser and cross-correlation operation, the scheme applies the state of the art of photon detection [14, 15].

However, for conventional OTDR, high sampling rate ADCs limit the development of the infinite backscatter technique. On the one hand, increasing the sampling rate of the ADC to a few gigahertz or more is difficult and expensive; on the other hand, the adoption of the high sampling rate ADC is inevitably accompanied by the problem of high-speed data cache and transmission, thereby increasing the complexity of the system design [16, 17]. In the PC-OTDR system, we can use a high-resolution time-to-digital converter (TDC) to replace the ADC in conventional OTDR. Based on the existing mature technology of TDC, it is easy to generate picosecond width time channels, or the effective sampling rate of terahertz can be obtained, so that the infinite backscatter technique can be further promoted with PC-OTDR [18]. In this letter, based on the principle of infinite backscatter, a scheme of infinite backscatter photon counting OTDR (IB-PC-OTDR) is proposed and experimentally demonstrated. Thanks to the high sensitivity of SPD and the high time resolution of TDC, IB-PC-OTDR offers higher dynamic range and better spatial resolution. In our proof of principle demonstration, we achieve a spatial resolution of 25.5 cm with 100 meters FUT, and a spatial resolution of 22.3 m for >50 km FUT. Specifically, for both cases, the spatial resolution keeps the same along the FUT, thus indicating that the spatial resolution with our proposed scheme is dispersion independent.

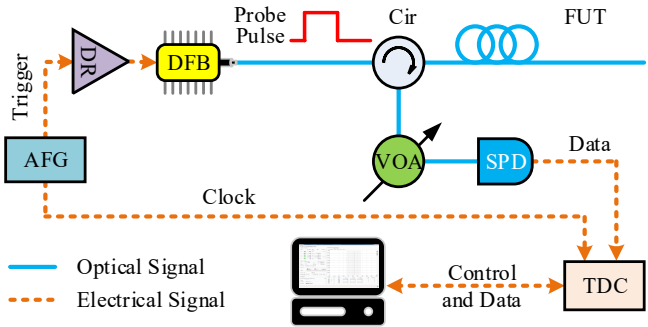

Fig. 1. Schematic of the IB-PC-OTDR. AFG: arbitrary function generator; DR: driver; DFB: distributed feed-back laser diode; FUT: fiber under test; VOA: variable optical attenuator; SPD: single photon detector; TDC: time-to-digital converter; Cir: circulator.

The schematic of IB-PC-OTDR is shown in Fig. 1. A distributed feed-back (DFB) laser diode operating at 1539.77 nm wavelength is used as the light source. It is directly driven by an arbitrary function generator (Tektronix, AFG3252) to periodically generate long light pulses. The probe pulses should satisfy the condition of only one traversing the fiber at a time. In this case, if $L$ is the length of FUT, the pulse width $\tau$ is $\tau \geq 2L/v_g$, and the pulse period $T$ is not less than $2\tau$, where $v_g$ is the group velocity of light in a fiber. The back-propagation signals are coupled into the third port of a circulator and then feed into a superconducting nanowire single photon detector (SNSPD) (Photon Technology Co., P-CS-6) [19]. The dark count rate of the SNSPD is 100 Hz with a detection efficiency of 58%, corresponding to a noise equivalent power of about $3.15 \times 10^{-18}$ W/$\sqrt{Hz}$ [20]. However, because the SNSPD has a dead time of 20 ns, it will cause the nonlinear error of photon counting when the incident photon rate is high [21]. In other words, the counting value is no longer proportional to the light intensity, and resulting in an inaccurate analysis of the loss curve. On the other hand, in order to obtain a high signal-to-noise ratio (SNR), the low count rate will lead to a long acquisition time [22]. Therefore, a variable optical attenuator (VOA) is placed in front of the SNSPD to adjust the count rate by controlling the power of back-propagation signals. Experimental results are obtained by using the output electrical pulses of AFG and SNSPD to trigger the TDC (ID Quantique, ID900), which works in histogram mode with a minimum time-bin width of 0.1 ns. Finally, the acquired data, containing the time delays between the laser pulse and the backscattered photon, are processed and displayed by a computer. We assume the width of time bin is $\Delta t$ and the number of signal photons in the $m$-th time bin is denoted $C[m]$, then the number of photons generated by back-propagation optical signals at the corresponding position ($0.5m \times \Delta t \times v_g$) is given by $C[m] - C[m-1]$. Due to using the histogram mode of the TDC, the sampling rate of the system is determined by the time-bin width, i.e. 0.1 ns in our experiment, the spatial resolution can be derived as $\Delta L = N \times \Delta t \times v_g$, where $N$ is a positive integer and represents the interval between the indices of time bins used to calculate the backscattered photon count. Obviously, the highest resolution can be achieved with $N = 1$.

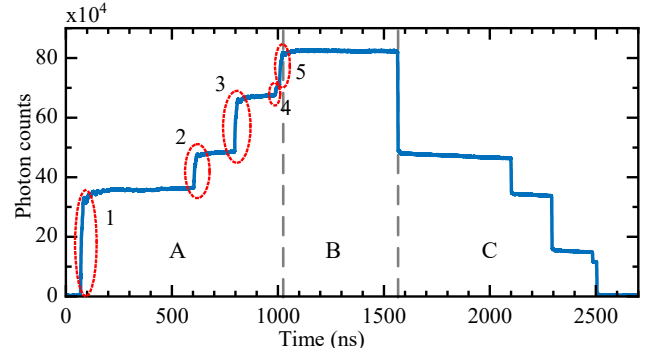

Fig. 2. Experimental data curve of FUT consisting of ~54 m, 20 m, 20 m and 2 m long fibers, obtained according to the output of TDC.

To evaluate the feasibility of IB-PC-OTDR, we show two experimental demonstrations using different lengths of FUT. In one demonstration, we use a piece of FUT consisting of a piece of standard single-mode fiber (SSMF, ~54 m), two pieces of 20 m long SSMF, and a piece of SSMF about 2 m, which connect to each other through connectors. The probe pulses are set to a pulse width of 1.5 μs with a period of 3 μs. Figure 2 is the experimentally obtained data, which reflects the whole process of the probe light from entering the FUT to leaving it. At the beginning of the curve, where the light has not yet entered the fiber, the photon counts mainly come from the dark count of the SNSPD with a rate of about 100 cps. Then, as the light propagating in the fiber, the curve ascends gradually and multiple discontinuities appear along it (see area A in Fig. 2). This is due to the existence of refractive index changing points along the FUT, resulting in Fresnel reflections, which are larger than the Rayleigh scattering. In our measurements, discontinuous points 1 to 4 are caused by optical fiber connectors, while the point 5 is caused by optical fiber end. Between adjoining discontinuous points, the measured photon counts increase with time. The value increased is proportional to the number of the backscattered photon at the corresponding position of the fiber. A flat curve (see area B in Fig, 2) after point 5 indicates that the FUT has been totally filled with probe light. Finally, as the probe light gradually outputs from the optical fiber, the photon counts of the back-propagation signals decrease over time (see area C in Fig, 2). The decreasing trend is the same as the first half of the curve.

The experimental results obtained from our setups contains two processes, i.e. probe pulse entering the fiber and quitting from it, corresponding to the area A and area C in Fig. 2, respectively. In principle, either rising or falling part should contain the attenuation and event information of the FUT, which can be used to perform the differential processing and draw the OTDR profile. However, for Rayleigh noise, the noise levels in different parts are different. Rayleigh noise due to the random change of Rayleigh scattering signal power is caused by the random phase of the source and the random variation of the polarization angle of the probe light in the fiber, and is proportional to the power of Rayleigh scattering signal [23]. Therefore, as the power of the back-propagation signals increase in the rising part, the level of the Rayleigh noise also increases. That is to say, with the same accumulating time, the OTDR profile analyzed by the falling part of the data curve has a higher SNR. This is consistent with the results obtained in our data processing. Please note that all the OTDR profiles in this work are obtained by using the data in the falling part for differential processing.

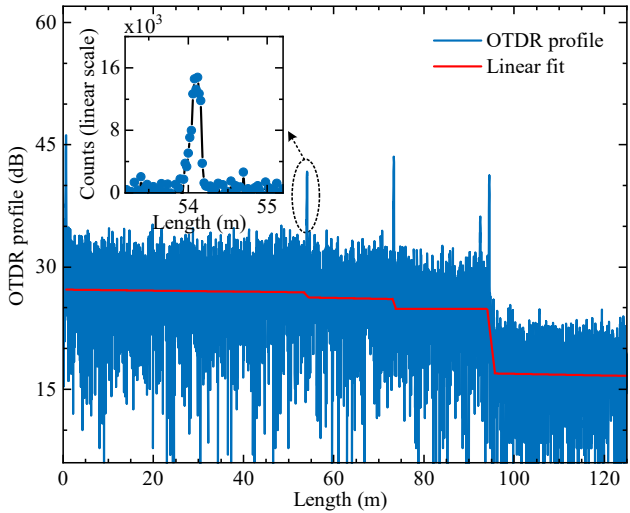

Fig. 3. OTDR trace for the FUT of about 100 m fiber. The inset shows the spatial resolution determination via the reflective peak analysis.

Figure 3 shows the result of the calculated OTDR trace. It is easy to see that the blue curve contains five reflection peaks, which is consistent with our setting in the experiment. The statistical noise is rather high, which can be significantly decreased by increasing the measurement time. For calculating the ultimate dynamic range, we performed a piecewise linear fit on the OTDR curve, as shown by the red line in the main part of Fig. 3 [24]. From the fitting curve, the dynamic range is about 10.3 dB, which is determined by the background-noise level and initial backscatter level. In the inset of Fig. 3, we present the analysis of the second reflective peak. In linear scale, the full width at $1/e^2$ criterion is used to determine the achievable spatial resolution, which is 25.5 cm. As the time-bin width of TDC is set to 200 ps, the highest spatial resolution should be $\Delta L = \Delta t \times v_g = 4$ cm. Such difference appears because the signal generated by the AFG for driving the DFB laser has a rising edge and a falling edge of 2.5 ns each, which is equivalent to a spatial distance of 25 cm. Hence, when the time-bin width is smaller than the pulse edge, the spatial resolution is mainly determined by the rising and falling edges of probe pulses in our proof of principle demonstration.

In the second demonstration, an SSMF of about 50 km long is used to generate the OTDR trace. The results are shown in Fig. 4. The probe pulses are set to a pulse width of 550 μs with a period of 1.2 milliseconds. Limited by the performance of TDC used in our demonstration, which has only 16384 histogram bins in total, the minimum time-bin width is set to 73.3 ns to ensure the complete acquisition of all backscatter signals in the 1.2 milliseconds cycle.

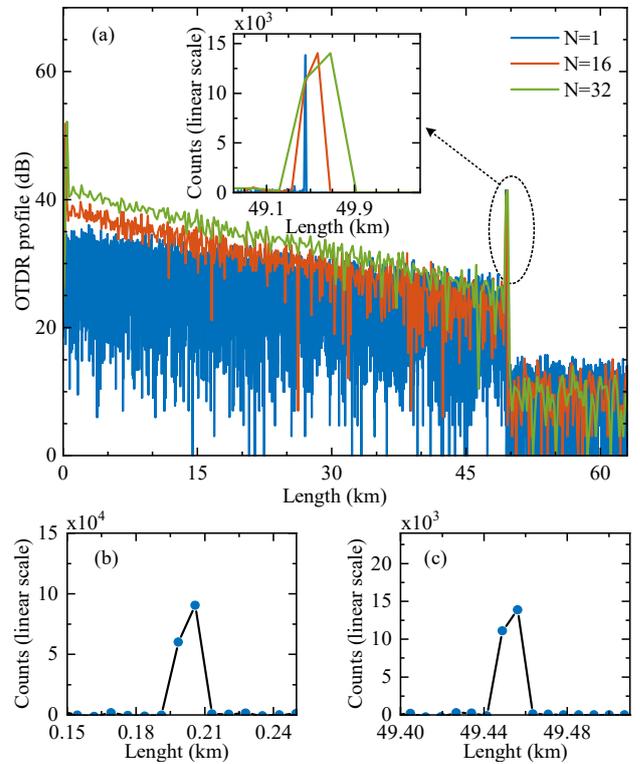

Fig. 4. OTDR traces for the link of about 50 km fiber. (a) The changes in dynamic range (main figure) and spatial resolution (inset) for various resolution configurations. (b) and (c) are analyses of the connectors at both ends of the fiber with $N$ equal to 1. $N$ represents the interval between the indices of time bins in differential processing.

Figure 4(a) shows the OTDR traces with three different $N$ values, corresponding to spatial resolutions of 22.26 m, 352.25 m, and 720.65 m, respectively. It is clear that the spatial resolution deteriorates as $N$ increases. On the other hand, the dynamic range increases with the increase of spatial resolution. The results show that the dynamic ranges are 27.16 dB (22.26 m), 39.06 dB (352.25 m) and 42.01 dB (720.65 m). When the spatial resolution is changed from 22.26 m to 352.25 m, the dynamic range is increased by 11.9 dB; when it is changed from 352.25 m to 720.65 m, it is increased by 2.95 dB. The improvements in dynamic range are in good accordance with the theoretical predictions, which should be 12.04 dB and 3.01 dB for our case (determined by the magnification of resolution $K$, i.e. $10\lg K$). A lower spatial resolution results in a higher dynamic range thus allowing longer distance measurements. In practical applications, once the data acquisition is completed, high dynamic range and high spatial resolution can be achieved simultaneously through changing indices of time bins in differential processing, i.e. the value of $N$.

To test the spatial resolution at different distances, we analyze the experimental data at both ends of the FUT. The results are shown in Fig. 4(b) and 4(c) with spatial resolutions of 22.25 m and 22.26 m for the front and the rear of FUT, respectively. It is clear from these data that along the fiber there is no reduction in spatial resolution, which is independent of the dispersion property of FUT. Note that the measured spatial resolution is still larger than the theoretical value of 14.66 m, determined by the time-bin width of 73.3 ns. Since the spatial distance of a time bin is much larger than the width of the reflection event, this theoretical spatial resolution can be obtained when a reflection event in the plotted OTDR curve corresponds to a single data point. However, in our demonstration, as shown in Fig. 4(b) and 4(c), there are two data points related to the same reflection event, resulting in a decrease in spatial resolution. This is because the laser pulse has wide edges and the rising edge is exactly in the two adjacent time bins. Fortunately, it can be distinctly optimized by adjusting the delay between the probe pulse and the clock or shortening the edges of laser pulse.

In conclusion, we propose and experimentally demonstrate a scheme for PC-OTDR based on infinite backscatter. The back-propagation OTDR signals along the fiber are obtained by differential analysis of a long laser pulse that illuminates or darkens the FUT gradually. This scheme drastically eliminates the need for high performance short-pulse lasers, and can develop PC-OTDR system with whose spatial resolution independent with dispersion property of the FUT. Our experimental results show that the IB-PC-OTDR achieves a spatial resolution of 25.5 cm with 10.3 dB dynamic range within a range of about 100 m and a spatial resolution of 22.3 m with 27.2 dB dynamic range within a range of about 50 km. The spatial resolution is affected by the resolution of TDC and the total number of time bins and the rising or falling edges of the probe laser pulse. Moreover, the dynamic range of the system improves with compromise in the spatial resolution. This means that our scheme would implement a high-dynamic and high-resolution measurement with only one data set.

Furthermore, the performance of our scheme can be improved by improving the experimental setups. For the state of the art, the time-bin width of TDC has been down to 203 fs, which means that the spatial resolution would be further shortened to ~40 μm [25]. Meanwhile the time-tagged time-resolved mode can be used to improve the test accuracy [26]. In addition, according to our latest work, i.e. externally time-gated technology, the performance of the IB-PC-OTDR can be further improved [27].

**Funding.** National Key R&D Program of China (2018YFA0307400, 2017YFA0304000); National Natural Science Foundation of China (NSFC) (61775025, 91836102, 61705033, 61405030, 61308041); Petro-China Innovation Foundation (2017D-5007-0603).

## References

1. E. Diamanti, C. Langrock, M. M. Fejer, Y. Yamamoto, and H. Takesue, "1.5 μm photon-counting optical time-domain reflectometry with a single-photon detector based on upconversion in a periodically poledlithium niobate waveguide," Opt. Lett. **31**, 727-729 (2006).
2. M. K. Barnoski, M. D. Rourke, S. M. Jensen, and R. T. Melville, "Optical time domain reflectometer," Appl. Opt. **16**, 2375-2379 (1977).
3. EXFO, "Spec sheets FTB-7600E Ultra-Long-Haul OTDR", https://www.exfo.com/umbraco/surface/file/download/?ni=10926&cn=en-US&pi=5453.
4. B. Li, Q. Zhou, R. Zhang, J. Li, H. Zhou, H. Li, Y. Ling, Y. Wang, G. Deng, Y. Wang, L. Shi, K. Qiu, and H. Song, "Cost-effective high-spatial-resolution photon-counting optical time-domain reflectometry at 850 nm," Appl. Opt. **57**, 8824-8828 (2018).
5. M. Legré, R. Thew, H. Zbinden, and N. Gisin, "High resolution optical time domain reflectometer based on 1.55μm up-conversion photon-counting module," Opt. Express **15**, 8237-8242 (2007).
6. A. L. Lacaita, P. A. Francese, S. D. Cova, and G. Riparmonti, "Single-photon optical-time-domain reflectometer at 1.3 μm with 5-cm resolution and high sensitivity," Opt. Lett. **18**, 1110-1112 (1993).
7. Q. Zhao, J. Hu, X. Zhang, L. Zhang, T. Jia, L. Kang, J. Chen, and P. Wu, "Photon-counting optical time-domain reflectometry with superconducting nanowire single-photon detectors," in *International Superconductive Electronics Conference*, (IEEE, 2013), pp. 1-3.
8. G. C. Amaral, J. D. Garcia, L. E. Y. Herrera, G. P. Temporão, P. J. Urban, and J. P. von der Weid, "Automatic Fault Detection in WDM-PON With Tunable Photon Counting OTDR," J. Lightwave Technol. **33**, 5025-5031 (2015).
9. S. Wang, X. Fan, and Z. He, "Ultrahigh Resolution Optical Reflectometry Based on Linear Optical Sampling Technique with Digital Dispersion Compensation," IEEE Photonics Journal **9**, 1-10 (2017).
10. L. Y. Herrera, G. C. Amaral, and J. P. von der Weid, "Ultra-High-Resolution Tunable PC-OTDR for PON Monitoring in Avionics," in *Optical Fiber Communication Conference*, OSA Technical Digest (online) (Optical Society of America, 2015), W2A.39.
11. J. Hu, Q. Zhao, X. Zhang, L. Zhang, X. Zhao, L. Kang, and P. Wu, "Photon-Counting Optical Time-Domain Reflectometry Using a Superconducting Nanowire Single-Photon Detector," J. Lightwave Technol. **30**, 2583-2588 (2012).
12. F. Calliari, L. E. Herrera, J. P. von der Weid, and G. C. Amaral, "High-dynamic and high-resolution automatic photon counting OTDR for optical fiber network monitoring," in *Proceedings of the 6th International Conference on Photonics, Optics and Laser Technology*, (2018), 82-90.
13. D. Devicharan, T. Zahnley, S. Dahl, A. Gurusami, and I. McClean, "Low optical power embedded OTDR," in *Optical Fiber Communication Conference*, OSA Technical Digest (online) (Optical Society of America, 2015), W4D.4.
14. Y. Wang, B. Wang, and A. Wang, "Chaotic Correlation Optical Time Domain Reflectometer Utilizing Laser Diode," IEEE Photonics Technology Letters **20**, 1636-1638 (2008).
15. X. Dong, A. Wang, J. Zhang, H. Han, T. Zhao, X. Liu, and Y. Wang, "Combined Attenuation and High-Resolution Fault Measurements Using Chaos-OTDR," IEEE Photonics Journal **7**, 1-6 (2015).
16. J.-X. Yang, "Design and Realization of Data Acquisition System for TOF-SIMS Based on High Speed ADC", master's thesis (Jilin University, 2018).
17. C.-G. Zong, "Time-of-flight mass spectrometer electronics reading method based on pulse front edge time measurement and amplitude correction algorithm", https://patents.google.com/patent/CN102789952A/zh.
18. W. Becker, *Advanced time-correlated single photon counting techniques* (Springer Science & Business Media, 2005), Vol. 81.
19. W. Zhang, L. You, H. Li, J. Huang, C. Lv, L. Zhang, X. Liu, J. Wu, Z. Wang, and X. Xie, "NbN superconducting nanowire single photon detector with efficiency over 90% at 1550 nm wavelength operational at compact cryocooler temperature," Science China Physics, Mechanics & Astronomy **60**, 120314 (2017).

20. J. S. Pelc, L. Ma, C. R. Phillips, Q. Zhang, C. Langrock, O. Slattery, X. Tang, and M. M. Fejer, "Long-wavelength-pumped up-conversion single-photon detector at 1550 nm: performance and noise analysis," Opt. Express **19**, 21445-21456 (2011).
21. Y. Hu, J. Li, M. Xia, D. Gao, and X. B. Zheng, "Measurement and Correction of Linearity of Photon Counters," Acta Photonica Sinica **45,** 0604001 (2016).
22. P. Eraerds, M. Legré, J. Zhang, H. Zbinden, and N. Gisin, "Photon Counting OTDR: Advantages and Limitations," J. Lightwave Technol. **28**, 952-964 (2010).
23. Y. Zhang, "Research on fiber link in-service monitoring method with large dynamic rang", master's thesis, (Beijing University of Posts and Telecommunications, 2018).
24. J. P. v. d. Weid, M. H. Souto, J. D. Garcia, and G. C. Amaral, "Adaptive Filter for Automatic Identification of Multiple Faults in a Noisy OTDR Profile," J Lightwave Technol **34**, 3418-3424 (2016).
25. Becker & Hickl , "SPC-150NXX Ultrafast TCSPC Module", https://www.becker-hickl.com/wp-content/uploads/2019/04/db-spc-150nxx-v01.pdf.
26. G.-L. Shentu, Q.-C. Sun, X. Jiang, X.-D. Wang, J. S. Pelc, M. M. Fejer, Q. Zhang, and J.-W. Pan, "217 km long distance photon-counting optical time-domain reflectometry based on ultra-low noise up-conversion single photon detector," Opt. Express **21**, 24674-24679 (2013).
27. B. Li, G. Deng, R. Zhang, Z. Ou, H. Zhou, Y. Ling, Y. Wang, Y. Wang, K. Qiu, H. Song, and Q. Zhou, "High Dynamic Range Externally Time-gated Photon Counting Optical Time-domain Reflectometry," J Lightwave Technol, DOI: 10.1109/JLT.2019.2941997.